\documentclass{sig-alternate-2013}
\newif\ifworkinprogress

\usepackage{amsmath}
\usepackage{comment}
\usepackage{graphicx}
\usepackage{balance}
\usepackage{dblfloatfix}
\usepackage{tikz}
\usetikzlibrary{positioning}
\usepackage{array}
\usetikzlibrary{arrows,%
                petri,%
                topaths}%
\usepackage{tkz-berge}
\usepackage{subfig}
\usepackage[font=bf,skip=\baselineskip]{caption}

\usepackage{amssymb}%
\usepackage{pifont}%
\usepackage{url}
\usepackage{multicol}
\usepackage{multirow}
\usepackage{array}
\usepackage{framed}
\usepackage{microtype}
\usepackage{balance}
\usepackage{xspace}
\usepackage{cleveref}
\usepackage{svg}
\usepackage{tabularx}
\usepackage[numbers,sort]{natbib}

\newcommand\Tstrut{\rule{0pt}{2.6ex}}         %
\newcommand\Bstrut{\rule[-0.9ex]{0pt}{0pt}}   %
\newcommand\tableSmallGap{\rule{0pt}{2ex}}   %

\newif\ifworkinprogressomments
\ifworkinprogress
    \newcommand{\mst}[1]{\textbf{\color{olive}/* #1 (mst) */}}
    \newcommand{\ddi}[1]{\textbf{\color{red}/* #1 (ddi) */}}
    \newcommand{\fle}[1]{\textbf{\color{red}/* #1 (fle) */}}
    \newcommand{\phsi}[1]{\textbf{\color{red}/* #1 (phsi) */}}
\else
    \newcommand{\mst}[1]{}
    \newcommand{\ddi}[1]{}
    \newcommand{\fle}[1]{}
    \newcommand{\phsi}[1]{}
\fi

\newcommand{\mytilde}{\raise.30ex\hbox{$\scriptstyle\mathtt{\sim}$}}

\newcommand{\footnoteurl}[1]{\footnote{\url{#1}}}
\newcommand{\para}[1]{\smallskip\noindent\textbf{#1}}

\makeatletter
\def\@copyrightspace{\@float{copyrightbox}[b]
\begin{center}
\setlength{\unitlength}{1pc}
\begin{picture}(20,4) %
\put(0,-0.95){\crnotice{\@toappear}}
\end{picture}
\end{center}
\end@float}
\makeatother

\permission{\copyright 2017 International World Wide Web Conference Committee \\ (IW3C2), published under Creative Commons CC BY 4.0 License.}
\conferenceinfo{WWW 2017,}{April 3--7, 2017, Perth, Australia.}
\copyrightetc{ACM \the\acmcopyr}
\crdata{978-1-4503-4913-0/17/04. \\
	http://dx.doi.org/10.1145/3038912.3052613 \\
	\includegraphics[scale=0.8]{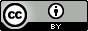}}

\begin{document}

\clubpenalty=10000 
\widowpenalty = 10000

\title{What Makes a Link Successful on Wikipedia?}

\numberofauthors{2}
\author{
   \alignauthor Dimitar Dimitrov\thanks{Both authors contributed equally to this work.}\\
     \affaddr{GESIS -- Leibniz Institute for the Social Sciences}\\
     \email{dimitar.dimitrov@gesis.org}\\
     \and
 \and
   \alignauthor Philipp Singer\footnotemark[1]\\
     \affaddr{GESIS -- Leibniz Institute for the Social Sciences}\\
       \affaddr{\& University of Koblenz-Landau} \\
     \email{philipp.singer@gesis.org}\\
     \and
   \alignauthor Florian Lemmerich\\
     \affaddr{GESIS -- Leibniz Institute for the Social Sciences}\\
     \affaddr{\& University of Koblenz-Landau}\\
          \email{florian.lemmerich@gesis.org}\\
     \and
\and
   \alignauthor Markus Strohmaier\\
     \affaddr{GESIS -- Leibniz Institute for the Social Sciences} \\
     \affaddr{\& University of Koblenz-Landau}\\
     \email{markus.strohmaier@gesis.org}\\
}

\maketitle

\begin{abstract}
While a plethora of hypertext links exist on the Web, only a small amount of them are regularly clicked. Starting from this observation, we set out to study large-scale click data from Wikipedia in order to understand \emph{what makes a link successful}. We systematically analyze effects of link properties on the popularity of links. By utilizing mixed-effects hurdle models supplemented with descriptive insights, we find evidence of user preference towards links leading to the periphery of the network, towards links leading to semantically similar articles, and towards links in the top and left-side of the screen.
We integrate these findings as Bayesian priors into a navigational Markov chain model and by doing so successfully improve the model fits. We further adapt and improve the well-known classic PageRank algorithm that assumes random navigation by accounting for observed navigational preferences of users in a weighted variation.
This work facilitates understanding navigational click behavior and thus can contribute to improving link structures and algorithms utilizing these structures.

\end{abstract}

\vspace{2mm}

\section{Introduction}

Even though links are omnipresent on the Web, only a minority of them get regularly clicked by humans.
For example, on Wikipedia only around $4\%$ of all existing links are clicked by visitors more frequently than 10 times within a month (cf. Section \ref{sec:discrepancy}). This phenomenon demonstrates the importance of a deeper understanding of human navigation behavior in order to provide efficient and effective navigational support for users. While a variety of navigational regularities, patterns and strategies have been identified in previous work \cite{west2012human,singer2015hyptrails,paranjape2016improving,lamprecht2016structure}, our research community still lacks a systematic understanding of \emph{factors that make a link successful in information networks}. Towards that end, we turn our attention to Wikipedia, one of the most popular information sources on today's Web.

\para{Problem and objectives.}
 In particular, we are interested in understanding \emph{the relationship between link properties and link popularity} as measured by \emph{large-scale transitional click data}. 
 Understanding what link features affect user click behavior can have implications for how we place links on Wikipedia, but can also have consequences for fundamental Web algorithms such as Google's PageRank that assume random navigation casting all links as equal and do not account for potential preferences towards specific links.

\begin{figure}[!t]
	\centering
	\includegraphics[width=0.45\textwidth]{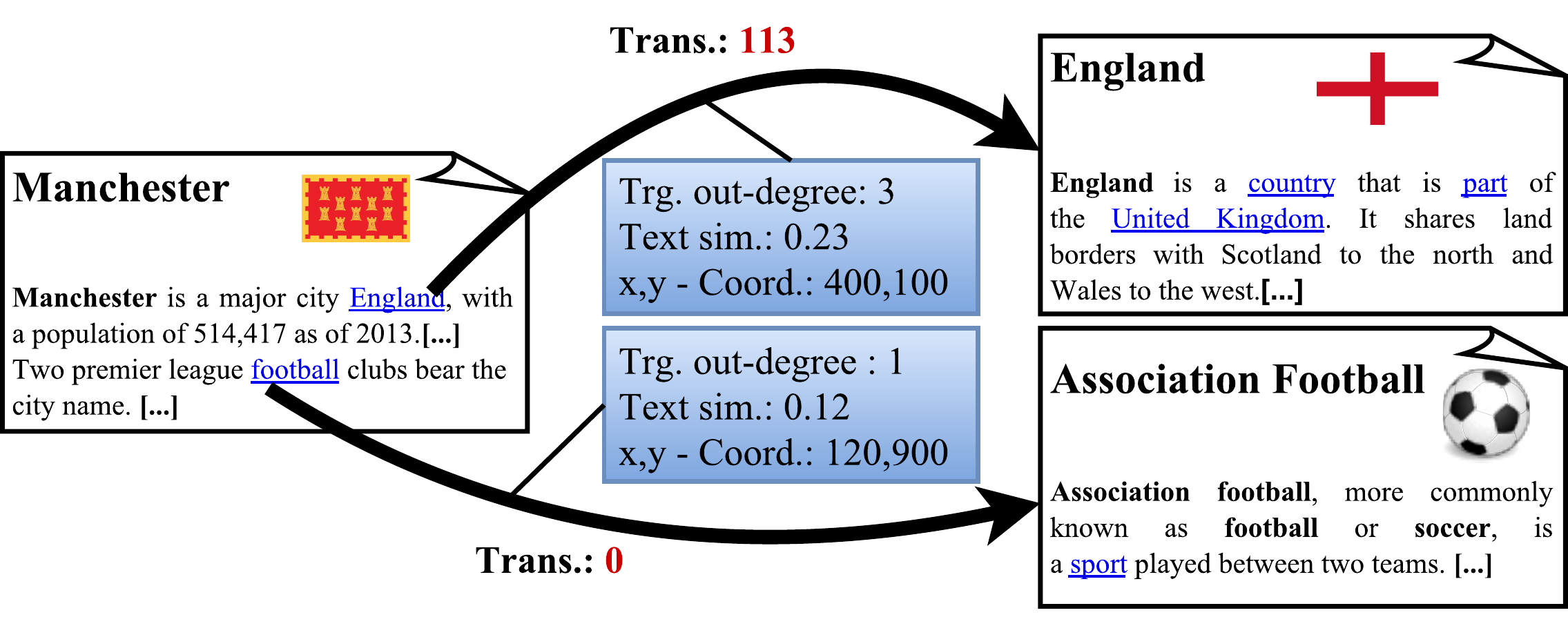}
	\caption{\textbf{Example.} Wikipedia pages are connected by links for which we can compute a variety of features that are categorized in \emph{network features} (e.g., target article's out-degree), \emph{semantic features} (e.g., text similarity between source and target articles), and \emph{visual features} (e.g., position of the link on the screen).
	This work aims at understanding what makes a link successful---i.e., which link properties best explain observed numbers of user transitions (shown on edges).}
	\vspace{-1em}
	\label{fig:manchester}
\end{figure}

\para{Materials, approach and methods.}
We utilize openly available large-scale datasets about the English Wikipedia: 
the network of internal links within Wikipedia articles and navigational data capturing user transitions between articles (cf. Section~\ref{sec:data}).
Motivated by previous work, we study three types of features that we compute from the link network and full texts of the articles: (i) \emph{structural features} (e.g., humans might prefer to navigate to central nodes in the information network) \cite{west2012human}, (ii) \emph{semantic features} (e.g., humans might prefer to navigate between semantically similar articles) \cite{singer2013}, and (iii) \emph{visual features} (e.g., humans might prefer links on the top of the screen) \cite{dimitrov2016visual}. 
For example, in Figure \ref{fig:manchester}, given the Wikipedia article ``Manchester'' and the out-going links to the articles ``England'' and ``Association Football'', we are interested in understanding which link features (e.g., the visual position of the link) can better predict the actual transitions (popularity) of a link.

Methodologically, we start by gaining descriptive insights into the success of links before we statistically model corresponding effects using mixed-effects hurdle models~\cite{pinheiro2006mixed,hu2011zero} allowing us to account for the heterogeneity in the data (e.g., links in the article ``USA'' might in general be more frequently used compared to links in the article ``Manchester'').
Subsequently, we integrate obtained insights into existing stochastic models of human navigation. To that end, we utilize the first-order Markov chain model---probably the most popular model for this task---in two separate analyses. 
First, we craft hypotheses about human navigation based on our insights and integrate these into a Bayesian inference process as priors; cf.~\cite{singer2015hyptrails}. Based on Bayesian model selection, this allows us to identify whether given hypotheses can improve the Markov chain model fit compared to an uninformed model; additionally, we can obtain a ranking of hypotheses allowing us to confirm our statistical results within a coherent research approach. Second, we aim at further utilizing these hypotheses to advance the classic well-known PageRank algorithm in a weighted variation relaxing the basic assumption of an uninformed random surfer; we evaluate the ranking of articles against actual page views.

\para{Contributions and findings.}
The main contributions of this paper are three-fold: (i)
We provide empirical evidence that Wikipedia users have a preference of choosing links leading to the periphery of underlying topological link network, that they prefer links leading to semantically similar articles, and that links positioned at the top and left-side of an article have a higher likelihood of being used. (ii) We integrate this evidence with first-order Markov chain models demonstrating an improvement of respective model fits by incorporating the assumed behavior of human navigation into the inference process (Section~\ref{sec:HypTrails}). (iii) Finally, we demonstrate the utility of these findings by adapting the well-known PageRank algorithm to better reflect human navigation behavior. Our enhancements lead to a significant improvement over the uninformed baseline algorithm that assumes random navigation by evaluating obtained ranking against actual article page views on Wikipedia (Section~\ref{sec:pagerank}). 
The methodological framework provided in this work can also easily be applied to other transitional click data apart from Wikipedia.

\section{Description of Data}
\label{sec:data}
This section describes the utilized datasets and extracted features.
We make an implementation of the data extraction process as well as a sample of the data publicly available online\footnote{\label{github}\url{https://github.com/trovdimi/wikilinks}}.

\subsection{Wikipedia link and transition data}

\para{Wikipedia data.}
In this work, we focus on all articles contained in the main namespace of the English Wikipedia as ex\-tra\-cted from the public XML dump from March, 2015\footnoteurl{https://archive.org/details/enwiki-20150304}.
To obtain authentically rendered pages, we retrieved the corresponding static HTML pages by using Wikimedia's public API \footnoteurl{https://www.mediawiki.org/wiki/API:Main_page}.
In contrast to using  readily available link dumps, this allowed also for considering links that are indirectly included in a page, e.g., by templates.
A tiny part (\mbox{$<0.01 \%$}) of articles could not be retrieved and had to be excluded from the analysis.
With this data, we created the Wikipedia link network $D_{wiki}$\xspace using articles as nodes and unique links as directed edges; $D_{wiki}$\xspace contains  \raise.30ex\hbox{$\scriptstyle\mathtt{\sim}$}$4.8$ million articles connected by \raise.30ex\hbox{$\scriptstyle\mathtt{\sim}$}$340$  million distinct links.

\para{Transition data.}
For measuring actual usage of links, we utilize openly available transition data from Wikipedia from February 2015~\cite{clickstream}. 
It contains aggregated page requests extracted from the server log for the English desktop version of Wikipedia in the form of \textit{(referrer, resource)} pairs, i.e., transitions, and their respective transition counts. 
The data has already been pre-processed to filter bots and web crawlers and transitions occurring less than 10 times, see \cite{clickstream} for details.
Since in this paper we are only interested in studying \emph{internal} navigation on Wikipedia, we excluded article requests from outside of Wikipedia (e.g., users arriving directly from search engines) and focus only on transitions corresponding to an internal link in our network $D_{wiki}$\xspace. %
These amount for about $1$ billion or $31\%$ of all page requests and cover \raise.30ex\hbox{$\scriptstyle\mathtt{\sim}$}$13.6$ million %
distinct links between 
\raise.30ex\hbox{$\scriptstyle\mathtt{\sim}$}$1.4$ million %
source and 
\raise.30ex\hbox{$\scriptstyle\mathtt{\sim}$}$2.1$ million %
target articles.
With this information we can weight edges in $D_{wiki}$\xspace by their transition counts to obtain an edge-weighted network $D_{trans_w}$\xspace. 
Additionally, we denote the sub-network of $D_{wiki}$\xspace that contains only edges (links) that we observe in the transition data---i.e., those links having a weight larger than 0 in $D_{trans_w}$\xspace---as $D_{trans}$\xspace.

\subsection{Link features}
\label{sec:linkfeatures}
For studying link success, we focus on three types of link features: network features, semantic similarity features, and visual features.

\para{Network features.}
Here, we capture the centrality of the source (\emph{src}) and target articles (\emph{trg}) of a specific link in the Wikipedia link network $D_{wiki}$\xspace considering the \emph{in-degree}, the \emph{out-degree}, the overall \emph{degree}, the \emph{pagerank}~\cite{page1999pagerank} and the \emph{k-core}~\cite{bader2003automated} measure.

\para{Semantic similarity features.}
Here, we detect the semantic similarity between the source  and target  article of a specific link. First, we computed a \emph{text similarity} score by utilizing a sub-linear scaled tf-idf weighted vector space model capturing word tokens of Wikipedia articles \cite{singer2015hyptrails}. For dimensionality reduction, we used sparse random projection retaining the pairwise distance between concepts with some small error.
The text similarity between two pages was then determined by the cosine similarity of corresponding vectors. We also computed a \emph{topic similarity} score with the cosine similarity between Wikipedia categories assigned to source and target.

\para{Visual features.}
The third group of features is based on the placement of links within the source  article, cf. also \cite{dimitrov2016visual}.
For that purpose, we rendered the HTML of each articles on a screen for a resolution of 1920 $\times$ 1080, a common resolution for desktop users.
This enables deriving the exact screen position of links in terms of their \textit{x/y-coordinates}. 
Based on visual position, CSS classes, and HTML tags, we additionally assigned each link to one of the following visual regions:\linebreak (1) \textit{lead}: all links in the first section of the article excluding infobox, (2) \textit{body}: all other links in the main text, (3) \textit{left-body}: links in the body that are displayed in the 10\% most left part of the screen, (4) \textit{right-body}: links in the remaining right part of the body,  (5) \textit{infobox}: all links in the infobox, (6) \textit{navbox}: all links in a table in the last section of the article placed by editors for facilitating navigation.
As links can occur multiple times on an article, we derived the visual position of a link based on its first occurrence on the screen. This goes hand-in-hand with our empirical results elaborated in Section~\ref{sec:results}. A more detailed discussion about this facet of our data is provided in Section~\ref{sec:discussion}.

\section{Focus of Attention}
\label{sec:discrepancy}
We start our analysis by investigating the focus of attention with respect to link usage in Wikipedia, i.e., we study how strongly user transitions are concentrated on the few most visited links. 

\begin{figure*}[ht!]
\centering
\subfloat[\textbf{Transitions counts}]{\includegraphics[width=0.33\textwidth]{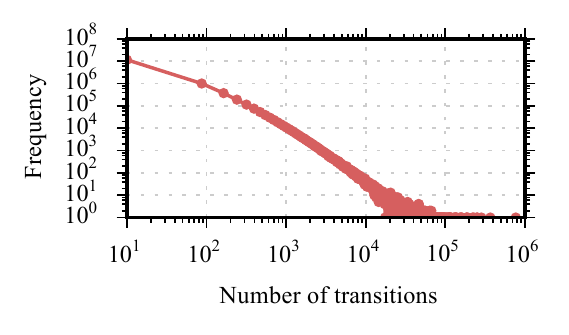}\label{fig:distributionOfTransitionCounts}}
\subfloat[\textbf{Out-degrees}]{\includegraphics[width=0.33\textwidth]{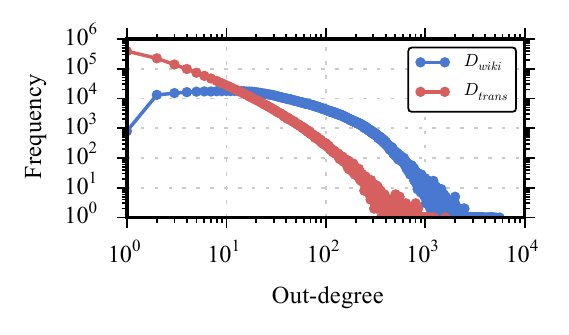}\label{fig:distributions_wiki_transitions_filtered}}
\subfloat[\textbf{Gini coefficients}]{\includegraphics[width=0.33\textwidth]{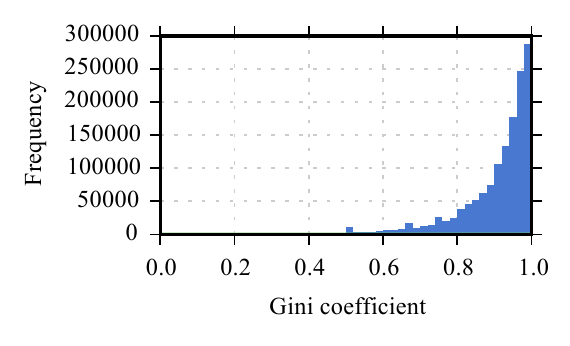}\label{fig:gini}}
\vspace{-0.5em}
\caption{\textbf{Focus of attention.} This figure illustrates the focus of users' attention towards specific links. In \protect\subref{fig:distributionOfTransitionCounts}, we see the frequency distribution of link transition counts. The long tail distribution shows that majority of links have low popularity. Figure \protect\subref{fig:distributions_wiki_transitions_filtered} compares the out-degree distributions of the Wikipedia link network $D_{wiki}$\xspace (blue) and the transition network $D_{trans}$\xspace (red) for nodes contained in both networks.	We can observe that only a few links are visited regularly on articles (peak at out-degree of one in $D_{trans}$\xspace) while more links are available (peak at out-degree of eleven in $D_{wiki}$\xspace); Figure \protect\subref{fig:gini} shows the histogram of Gini coefficients calculated on article vectors of link transition counts revealing clear inequality of link popularity within articles.}
\vspace{-1.5em}
\label{fig:focusOfAttention}
\end{figure*}

\para{Distribution of transition counts.}
As a first key statistic, we observe that of \raise.30ex\hbox{$\scriptstyle\mathtt{\sim}$}$340$ million links in $D_{wiki}$\xspace only about $13.6$ million links (or about 4\% of all links) are regularly used, i.e., are visited at least 10 times within a month.
However, even when considering only these 4\% regularly used links, we can observe that users focus heavily on a comparatively small set of links: about 50\% of all transitions stem from only about $600,000$ links.
A frequency plot of the distribution, see Figure~\ref{fig:focusOfAttention}\subref{fig:distributionOfTransitionCounts}, reveals a long-tailed distribution, where most of the links receive very low amounts of transitions.%

\para{Out-degree distribution. }
A similar effect can be observed at the level of individual articles.
We compare the out-degree distributions of nodes in $D_{wiki}$\xspace and $D_{trans}$\xspace, i.e., we compare the number of available links on articles in $D_{wiki}$\xspace with the number of regularly used links in $D_{trans}$\xspace. 
For a fair comparison, we consider only the out-degrees of nodes that occur in both networks. The results in Figure~\ref{fig:focusOfAttention}\subref{fig:distributions_wiki_transitions_filtered} reveal clear differences. 
The out-degree distribution of $D_{wiki}$\xspace reaches a maximum frequency for articles having eleven out-links indicating a decent amount of links to choose from. 
However, the out-degree distribution of $D_{trans}$\xspace clearly reveals that for most articles, users only visit a few articles regularly with the maximum frequency of only a single link.
Formal comparison of both distributions with respect to (truncated) power-law, exponential, and log-normal distributions by their goodness of fit reveals that for $D_{trans}$\xspace the out-degree distribution resembles a truncated power-law most closely, while for $D_{wiki}$\xspace a log-normal distribution is preferred.

\para{Gini coefficient of transition counts. }
The previous analyses have suggested a discrepancy between the link network $D_{wiki}$\xspace and actually transitioned links in $D_{trans}$\xspace and $D_{trans_w}$\xspace. 
The \emph{Gini coefficient} is a measure of statistical dispersion allowing us to quantify the inequality of link popularity within articles. We calculate the Gini coefficient for each article based on a vector containing the transition counts for all links (including zeros) in this article from $D_{trans_w}$\xspace. A Gini coefficient of 0 indicates complete equality, while a result of 1 would mean that only a single link gets the whole attention. We visualize the histogram of all coefficients for each article in Figure~\ref{fig:focusOfAttention}\subref{fig:gini}; results indicate that there is a clear inequality of link popularity within articles, i.e., attention focuses on just a few links.

\para{Summary.} In this section, we have presented empirical evidence that on Wikipedia user attention focuses on a small set of links while navigating. This hints towards a clear \emph{preference towards a few links} and demonstrates the importance of finding indicators of link success, on which we focus in the remainder of this paper.

\section{What makes a link successful?}
\label{sec:success}
Motivated by the discrepancy between link presence and link usage, we
now study link success. That is, as illustrated in Figure~\ref{fig:manchester}, given a Wikipedia article, we aim at identifying effects of link features (network features, semantic features, or visual features) on counts of transitions to the respective target article.
This allows us to answer questions such as: ``Are links leading to peripheral articles more popular and thus, more successful?''
We propose a methodology in Section~\ref{sec:method} and discuss results in Section~\ref{sec:results}.

\vspace{2mm}
\subsection{Methodology}
\label{sec:method}

For studying effects of link features on the success of links, we employ  descriptive statistics and \emph{hurdle mixed-effects regression models}; we outline both approaches next.

\para{Descriptive analysis.}
We start our analysis with a  descriptive overview on the effects of our features on transition counts.
For an intuitive picture, we investigate boolean expressions on our features and determine for each feature the number of links, the number of transitions, and the average number of transitions per link.
We focus on easy-to-interpret expressions: for network features, we compare the centrality of the target node with the centrality of the source node. 
For semantic similarity features, we compare the similarity between source and target article of the link with the median of the respective values for all links within the same source article.
Finally, for the visual features, we divide the screen position in three areas for each dimension, that is the left, the middle, and the right third of the screen, and links in the top-half, the bottom-half and links that are only visible after vertical scrolling (assuming a resolution of 1920 $\times$ 1080) respectively.
Furthermore, we use the position of the link in a specific part of the article such as the lead or the infobox.

\para{Mixed-effects hurdle model.}
Additionally, we aim at statistically modeling the effects of all link features on link success given a certain start node. 
Our setting (cf. Figure~\ref{fig:manchester}) implies that individual links represent data points with the transition counts being the outcome to predict, and the respective link features the effects at interest. Yet, links in the dataset are not independent of each other, but rather nested within Wikipedia articles. 
Heterogeneous effects might be present as e.g., apart from generally different amounts of total transitions, the article of ``Manchester'' might pose different navigational behavior than the article of ``USA''.
Pooling all links into a single dataset for ordinary regression can lead to biased results; consequently we resort to \emph{mixed-effects models} (also known as multilevel models or hierarchical models), allowing us to account for the nested structure---specifically the resulting random variations---between links (level 1) and Wikipedia pages (level 2).
We define our mixed-effects model following \cite{hox2010multilevel}; for a broader introduction we point to \cite{pinheiro2006mixed}:
\begin{equation}
    Y_{i,j} = \gamma_{0,0} + \gamma_{1,0}X_{1,i,j}+u_{0,j}+u_{1,j}X_{1,i,j}+e_{i,j}
\end{equation}
Here, $Y_{i,j}$ refers to the transition count of a link $i$ (level 1) on a Wikipedia page $j$ (level 2); $\gamma_{0,0}$ refers to the overall intercept and $\gamma_{1,0}$ to the overall regression coefficient for the level 1 predictor $X_{1,i,j}$, e.g., the degree of the target node. 
These are the fixed effects.
The remainder refers to the random effects: $u_{0,j}$ is the error component of the intercept allowing the group intercepts to differ, $u_{1,j}$ refers to the error component of the slope allowing the group slopes to differ, and $e_{i,j}$ is the classic random error of predictions.

To study the effects of network, semantic similarity and visual features, we model each feature separately.
As shown in Section~\ref{sec:discrepancy}, only 4\% of all links have been used at least 10 times, resulting in a large amount of zeros in the data outcomes.
To account for that, we employ \textit{hurdle models} \cite{hu2011zero} in a two-step approach.
(i) First, we model the binary data in $D_{trans}$\xspace; i.e., the link values are set to $1$ iff at least $10$ transitions for these links have been observed, and to $0$ otherwise. We then model this data with a mixed effect binomial logistic regression and make inference on the fit. 
(ii) In a second step, we only take the data from $D_{trans_w}$\xspace capturing those links that have been clicked and their respective transition counts. 
As our outcome variable is generated by count data, 
basic assumptions for \emph{linear} mixed-effects models are violated.
Consequently, we resort to zero-truncated negative binomial mixed-effect models that  do not allow the outcome to be zero (as all outcome counts are at least $10$).
The choice of model also enables modeling the presence of overdispersion in the data. 
For comparison, we also fitted Poisson mixed-effects models with an observation-level random effect that allows for overdispersion \cite{harrison2014using} but could not find core differences in inference.
Thus, we overall build two models for each feature: one to model \emph{if} a link has been used at least 10 times in a month, and a second to model the transition \emph{counts} for these regularly used links.

 We fit our models using the ``Template Model Builder'' framework \cite{kristensen2015tmb}
 that evaluates and maximizes the Laplace approximation of the marginal likelihood utilizing automatic differentiation of the joint likelihood; we use the R package ``glmmTMB''\footnoteurl{https://github.com/glmmTMB/glmmTMB}. 
 For improving convergence, we scale non-binary features by subtracting the mean and dividing by their standard deviation.
 For judging significance of individual effects, we follow an incremental model comparison approach \cite{bates2014lme4}.
 That is, we compare the model including the effect of interest with a restricted model that neglects the effect using a  \emph{likelihood ratio chi-square test}. Results could be confirmed by model comparison via AIC and BIC.

\begin{table}[b!]
\vspace{-1em}
\setlength\tabcolsep{2px}
\fontsize{6}{7.2}\selectfont
  \centering
  \caption{\textbf{Descriptive results.} This table provides a descriptive overview on link success; it shows for each boolean feature the number of links, the overall number of transitions these links accumulated, and the average transition count per link. Results show that links leading to the periphery of the network, to semantically similar articles, and links positioned in the lead and left area of an article accumulate relatively more transitions than opposite links.} 
\vspace{-0.5em}
\begin{tabularx}{\columnwidth}{|X|rrr|}
    \hline
    Feature & \# Links & \# Transitions &
        \begin{tabular}[x]{@{}r@{}}Mean Trans.\\per Link\end{tabular}
        \Tstrut\Bstrut \\
    \hline
    \Tstrut\Bstrut
    trg\_degree $<$ src\_degree & 462,147 & 3,944,747 & 8.54 \\[1mm]
    trg\_degree $\geq$ src\_degree & 566,557 & 2,741,834 & 4.84 \\[1mm]
    trg\_in\_degree $<$ src\_in\_degree & 412,653 & 3,831,100 & 9.28 \\[1mm]
    trg\_in\_degree $\geq$ src\_in\_degree & 616,051 & 2,855,481 & 4.64 \\[1mm]
    trg\_out\_degree $<$ src\_out\_degree & 545,980 & 3,912,604 & 7.17 \\[1mm]
    trg\_out\_degree $\geq$ src\_out\_degree & 482,724 & 2,773,977 & 5.75 \\[1mm]
    trg\_kcore $<$ src\_kcore & 283,668 & 3,477,828 & 12.26 \\[1mm]
    trg\_kcore $\geq$ src\_kcore & 745,036 & 3,208,753 & 4.31 \\[1mm]
    trg\_page\_rank $<$ src\_page\_rank & 414,124 & 3,833,596 & 9.26 \\[1mm]
    trg\_page\_rank $\geq$ src\_page\_rank & 614,580 & 2,852,985 & 4.64 \\[1mm]
    \hline
    text\_sim $>$ median of article & 511,871 & 4,335,445 & 8.47 \\[1mm]
    text\_sim $\leq$ median of article & 516,833 & 2,351,136 & 4.55 \\[1mm]
    topic\_sim $>$ median of article & 392,955 & 3,246,766 & 8.26 \\[1mm]
    topic\_sim $\leq$ median of article & 635,749 & 3,439,815 & 5.41 \\[1mm]
    \hline
    position = lead & 93,546 & 2,489,342 & 26.61 \\[1mm]
    position = body & 373,407 & 3,442,017 & 9.22 \\[1mm]
    position = left-body & 64,320 & 934,766 & 14.53 \\[1mm]
    position = right-body & 309,087 & 2,507,251 & 8.11 \\[1mm]
    position = navbox & 502,079 & 99,498 & 0.2 \\[1mm]
    position = infobox & 59,672 & 655,724 & 10.99 \\[1mm]
    screen\_x\_coord: left third & 211,549 & 2,358,899 & 11.15 \\[1mm]
    screen\_x\_coord: middle third & 291,260 & 1,422,183 & 4.88 \\[1mm]
    screen\_x\_coord: right third & 525,847 & 2,905,499 & 5.53 \\[1mm]
    screen\_y\_coord: top half & 221,980 & 3,991,912 & 17.98 \\[1mm]
    screen\_y\_coord: bottom half & 174,500 & 998,287 & 5.72 \\[1mm]
    screen\_y\_coord: scroll needed & 632,176 & 1,696,382 & 2.68 \\[1mm]
    \hline
    \Tstrut\Bstrut Overall & 1,028,704 & 6,686,581 & 6.5 \\[1mm]
    \hline
\end{tabularx}%
 \label{tab:empiricalOverviewFeatures}
\end{table}

\begin{table*}[t!]
\centering
\caption{\textbf{Mixed-effects hurdle modeling results.} This table presents the results for the mixed-effects hurdle models based on a basic model specification of: $\text{transition}_{i,j} = \gamma_{0,0} + \gamma_{1,0}\text{feature}_{1,i,j}  +  u_{0,j} + u_{1,j}\text{ feature}_{1,i,j} + e_{i,j}$. 
For each feature (first column) and a given transformation (second column), results for both parts of the hurdle model are reported. The binomial models the binary outcome whether a link has been used at all, and the zero-truncated negative-binomial models the transition counts of regularly used links.
For each model, we report the fixed effect coefficient of the fitted model, the likelihood ratio test statistic compared to a reduced nested model without the fixed effect, and the significance according to $\chi^2$-test.
Overall, the results  confirm the descriptive results of Table~\ref{tab:empiricalOverviewFeatures} suggesting a preference of links leading to the periphery of the network, links targeting semantically similar articles, and links being positioned at the beginning and left-side of articles. 
}
\vspace{-0.5em}
\begin{tabular}{|l|c|ccc|ccc|}
\hline
& & \multicolumn{3}{c|}{Binomial Model} & \multicolumn{3}{c|}{Zero-truncated NB Model}\Tstrut\Bstrut \\ %
Feature & Transformation  & Fixed effect & LRT & Pr($>$Chi) & Fixed effect & LRT & Pr($>$Chi)  \Tstrut\Bstrut \\ \hline

\Tstrut\Bstrut trg\_degree  &  scale  &  -9.0704  &  9441  &    $<$ 1.0 e-300 ***  &  -0.1353  &  174  &  7.98 e-40 *** \\ 
trg\_in\_degree  &  scale  &  -9.2925  &  9456  &    $<$ 1.0 e-300 ***  &  -0.1341  &  185  &  4.2 e-42 *** \\ 
trg\_out\_degree  &  scale  &  -0.6425  &  4052  &    $<$ 1.0 e-300 ***  &  -0.0375  &  32  &  1.6 e-08 *** \\ 
trg\_kcore  &  scale  &  -9.4873  &  9470  &    $<$ 1.0 e-300 ***  &  -0.1296  &  180  &  4.5 e-41 *** \\ 
trg\_pagerank  &  scale  &  -8.8103  &  9384  &    $<$ 1.0 e-300 ***  &  -0.1457  &  196  &  1.7 e-44 *** \\ 
\tableSmallGap text\_sim  &  scale  &  0.2944  &  617  &  3.4 e-136 ***  &  0.238  &  1179  &  1.9 e-258 *** \\  
topic\_sim  &  scale  &  0.2052  &  482  &  8.6 e-107 ***  &  0.1212  &  401  &  3.7 e-89 *** \\  
\tableSmallGap position = lead  &  none  &  1.9682  &  6146  &    $<$ 1.0 e-300 ***  &  0.2858  &  498  &  2.7 e-110 *** \\  
position = body  &  none  &  0.2351  &  93  &  5.9 e-22 ***  &  -0.4581  &  1478  &     $<$ 1.0 e-300 *** \\  
position = left-body  &  none  &  1.106  &  739  &  9.4 e-163 ***  &  -0.1671  &  123  &  1.7 e-28 *** \\  
position = right-body  &  none  &  -0.3631  &  231  &  4.4 e-52 ***  &  -0.4389  &  1236  &  1.0 e-270 *** \\ 
position = infobox  &  none  &  -0.1955  &  25  &  7.2 e-07 ***  &  0.1365  &  37  &  1.4 e-09 *** \\  
position = navbox  &  none  &  -6.3237  &  9226  &    $<$ 1.0 e-300 ***  &  -0.9963  &  688  &  1.5 e-151 *** \\ 
screen\_x\_coord  &  scale  &  -0.2974  &  1081  &  4.4 e-237 ***  &  0.0225  &  16  &  6.0 e-05 *** \\ 
screen\_y\_coord  &  scale  &  -4.258  &  10588  &    $<$ 1.0 e-300 ***  &  -0.4004  &  419  &  3.987 e-93 *** \\ \hline

\end{tabular}
\label{tab:results}
\end{table*}

 \subsection{Results}
\label{sec:results}

For computational reasons, we applied the methodology to a random sample of $10,000$ 
articles containing at least one out-going link in the transition data $D_{trans}$\xspace considering all links and their features. 
The sample contains $1,028,704$ links with $6,686,581$ transitions.
We made sure that the sample captures similar power-law degree distributions as the overall dataset (cf. Section~\ref{sec:discrepancy}); multiple sampling iterations did not change the results. For reproducibility reasons, we make the full sample available\footnoteurl{https://github.com/trovdimi/wikilinks/raw/master/sample.csv.gz}. %
We present the descriptive results in Table~\ref{tab:empiricalOverviewFeatures} and the mixed-effects model results in Table~\ref{tab:results}. For the statistical models, we only report fixed effects at interest but make the full results available in additional material\footnoteurl{https://github.com/trovdimi/wikilinks/tree/master/notebooks}. %
We discuss results for our three major feature groups next.

\para{Network features.}
The results in Table~\ref{tab:empiricalOverviewFeatures} highlight that there appears to be a preference of users to choose links 
leading to target Wikipedia articles that have a smaller in-degree, out-degree and degree compared to the source article. This is specifically imminent when controlling for the presence of links as the last column indicates. Similar behavior can be seen for the k-core and pagerank of nodes. 
Our mixed-effects model results shown in Table~\ref{tab:results} statistically confirm this observation. The coefficients for both parts of the hurdle model (binomial and zero-truncated negative binomial models) show negative signs for all network features at hand meaning that the higher those features are (i.e., the more central the target nodes are) the less likely it is that those links are clicked. Interestingly, the effect sizes and model fits (LRT) are quite similar for all features except the out-degree which can be explained by the fact that in general the out-degree is not a good indicator regarding the centrality of a node in the network as it just depends on the number of links present on a given page.
Overall, the results suggest a preference for users to navigate to peripheral nodes in the network in contrast to navigating towards the core of the network.
A potential explanation of this behavior could be that pages located in the core of the network are more general, whereas pages located in the periphery of the network are more specific and thus possibly more interesting to the user on average. Moreover, more general Wikipedia pages are likely more often returned as results to search engine queries, making them the entry point to Wikipedia. Then, with the next click within Wikipedia, the user narrows down her information need. Such a behavior---looking up specific facts or information---would be in-line with a common use of encyclopedias in general. 

\para{Semantic similarity features.}
The results in Table~\ref{tab:empiricalOverviewFeatures} and Table~\ref{tab:results} indicate that links connecting two semantically related Wikipedia articles---both for text and topic similarity---are more likely to be transitioned compared to those links that indicate less semantic similarity. This is imminent from the higher ratio of transitions to high similarity target articles, and specifically, by the positive mixed-effects model coefficients for the hurdle models at hand. Overall, the results suggest a preference towards navigating semantically similar articles supporting
previous research \cite{west2012human,singer2013,singer2015hyptrails}.

\para{Visual features.}
The descriptive results in Table~\ref{tab:empiricalOverviewFeatures} indicate user preference towards choosing links on the left side of the screen and specifically links that are at the beginning of a Wikipedia article. 
The model results in Table~\ref{tab:results}  confirm these observations. The positional lead feature has strongly positive coefficients in both the binomial and zero-truncated negative binomial model indicating that links in the lead are more frequently used. 
Additionally, we can see that the right-body feature (links in the right 90\% body region) indicates negative coefficients while the left-body feature (links in the left 10\% body region) has a positive coefficient for the binomial model. The models for the infobox are contradicting each other, but overall show much less model improvements over the baseline model compared to our other visual feature models (lower LRT). Furthermore, we can clearly see that the navbox feature has a strong negative coefficient suggesting that links positioned at the bottom of the page in the navigation boxes are only seldomly visited. The coordinate features summarize the more fine-grained results in a broader level indicating that there is a preference towards choosing links that are at the beginning and also on the left side of the screen.
The results presented are by and large confirming previous findings on the F-shaped way of information consumption \cite{dimitrov2016visual, nielsen2016Fshaped, buscher2009you}. 

\para{Summary.} Our results suggest three indicators having an effect on the success of links on Wikipedia. First, humans appear to have a \textit{preference towards choosing links leading to the periphery} of the underlying topological link network. Second, \textit{links connecting semantically related articles tend to be clicked more frequently} than those connecting semantically unrelated articles. Third, the visual position of a link appears to have an impact of the link's success. Links that are located on the top of the screen (i.e., lead of an article) and on the left of the screen are preferred by humans navigating Wikipedia suggesting a \textit{visual top-left preference}.

\section{Integration in Markov Models}
\label{sec:markov}

While the results presented in Section~\ref{sec:success} provide evidence of factors effecting the success of links on Wikipedia, it is an open question how we can integrate these findings into existing models of human navigation.
To that end, we investigate the influence of link features on models of human navigation in this section.
We focus on first-order Markov chains as the most wide-spread models for this task \cite{borges2000data,brin1998anatomy,singer2014detecting}.
First-order Markov chains are stochastic systems that assume a memoryless generative process, i.e., the probability of the next state (Wikipedia article) depends only on the current state.
We utilize these models in two separate analyses.
First, in Section~\ref{sec:HypTrails}, we craft hypotheses about human navigation based on our insights of Section~\ref{sec:results} and incorporate these into a Bayesian inference process. This allows us to (a) identify whether given hypotheses can better explain observed transition behavior compared to a uniform structural baseline hypothesis and thus, potentially improve the Markov chain model fit, and (b) the model comparison provides an additional ranking of given hypotheses helping in confirming our previous results within a coherent research approach.
Second, in Section~\ref{sec:pagerank}, 
we aim at utilizing these insights to improve the classic PageRank algorithm in a weighted variation and validate its precision against the observed visit counts of Wikipedia articles.

\subsection{Bayesian integration}
\label{sec:HypTrails}
Next, we explore the impact of integrating our findings in Markov chain navigation models; for that task we resort to a recently proposed Bayesian approach called HypTrails~\cite{singer2015hyptrails}. 

\para{Methodology.}
We aim to integrate hypotheses about navigational behavior, which we form based on our results obtained in Section~\ref{sec:results}, as prior knowledge into the inference process for Markov chains. HypTrails~\cite{singer2015hyptrails} provides a coherent framework for this task. 
First, one can express a hypothesis, i.e., beliefs in transition probabilities of a Markov chain, as a matrix $M = (m_{ij})$, where a higher value expresses higher belief in a transition between articles $i$ and $j$.
Then, these hypotheses matrices are incorporated as Dirichlet priors into the Bayesian model inference procedure. 
Due to the sensitivity of a Bayesian model to the priors on its parameters, more plausible hypotheses---i.e., hypotheses that are in line with the observed data---induce higher marginal likelihood (evidence) values used for model comparison.
Marginal likelihoods can be compared along different values of a parameter $\kappa$ expressing the strength of belief in a hypothesis. Higher values reflect a higher belief in given hypothesis; 
i.e., a stronger probability mass concentration of the Dirichlet prior.
Formally, model comparison is done by calculating a \emph{Bayes factor} \cite{kass1995bayes}, which is the ratio of the  marginal likelihoods for two hypotheses $H_1$ and $H_2$; if positive, the first hypothesis is judged as more plausible---strength of Bayes factors can be checked in an interpretation table of Kass and Raftery \cite{kass1995bayes}.

Comparing the marginal likelihood of a specific hypothesis (e.g., that humans prefer to navigate to the periphery of the network) with a structural baseline hypothesis (i.e., random preference of out-links) allows us to detect whether integrating given hypothesis into the inference process enhances the general model fit. Additionally, we can compare hypotheses with each other to obtain a ranking of hypotheses.
For more methodological details, we point the reader to ~\cite{singer2015hyptrails}.
We describe how we construct hypotheses for our data next.

\para{Hypotheses.}
We formalize and compare several hypotheses based on our main findings of Section~\ref{sec:success} providing evidence of feature effects on link success. 
As mentioned, hypotheses are expressed as matrices $M = (m_{ij})$ indicating belief in transition probabilities. 
For finding good formalizations of these hypotheses, we experimented with different scaling variations such as logarithmic, square-root, and exponential feature scaling, but only report on the best formalization for each of the three feature groups due to limited space.

As a baseline, we use the \emph{structural hypothesis} expressing the belief that someone navigating in the Wikipedia network chooses any out-link of an article randomly.
Here, we set $m^{structure}_{ij} = 1$ if there is a link from article $i$ to article $j$ and $m^{structure}_{ij}= 0$ otherwise. If other hypotheses do not reveal higher marginal likelihoods than this baseline hypothesis (i.e., positive Bayes factors), they indicate no improvements to the model fit---i.e., they do not capture the actual human navigation behavior well.

For the network features, we focus on the \emph{k-core} having shown promising results in Section~\ref{sec:success}.
Since we want to formulate a hypothesis that expresses higher beliefs in transitions to articles with lower k-cores (i.e., the periphery of the network), we set the entries of the hypothesis matrix to $m^{kcore}_{ij}=\frac{1}{\sqrt{trg\_kcore}}$.
Regarding semantic similarity, we use the \emph{text\_sim} feature as it is, i.e., $m^{text\_sim}_{ij}= text\_sim$. 
This reflects a belief that transitions to similar articles (w.r.t. to the text similarity) are more likely.
For visual features, we concentrate on the \emph{position} feature. 
Our visual hypothesis states that links that are in the lead, in the most left part of the body, or in the infobox are more likely to be visited than the others (as the results of Section~\ref{sec:success} suggest).
Thus, we express this hypothesis as $m^{visual}_{ij} = 1$ in these cases, and $m^{visual}_{ij} = 0$ otherwise.
For these hypotheses, we add the structural baseline hypothesis for smoothing in order to guarantee a minimum belief in all transitions to linked articles.

In addition to these individual hypotheses, we also include combinations of them in our comparison.
In this direction, we form the hypotheses \emph{kcore+text\_sim}, \emph{kcore+visual}, \emph{text\_sim+visual}, and \emph{kcore+text\_sim+visual} by adding the respective matrices element-wise to each other---here, no smoothing is necessary.

\begin{figure}[!t]

	\centering
	\includegraphics[width=\linewidth]{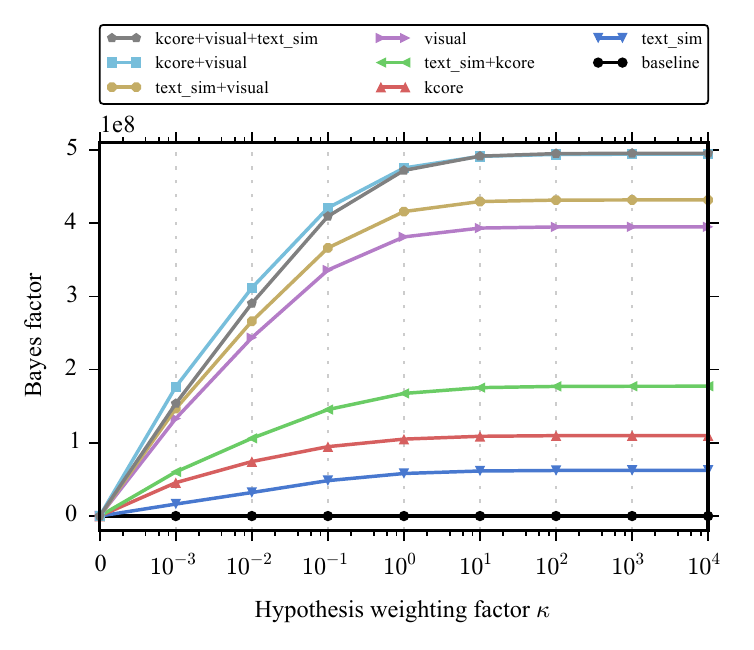}
	\vspace{-2.5em}
	\caption{\textbf{Bayesian integration results.} This figure reports the results for our experiments on integrating navigational hypotheses as Bayesian priors in Markov chain models. The x-axis depicts
	different hypothesis weighting factors $\kappa$---higher values reflect stronger belief in a given hypothesis. The y-axis denotes the Bayes factor for a given hypothesis (different lines); the Bayes factor compares the fit of the respective hypothesis model to the fit of the baseline, i.e., the structural hypothesis. Higher Bayes factors reflect higher plausibility of given hypothesis; all Bayes factors in this plot are strongly decisive. %
	All hypotheses improve the fit of the Markov chain model when integrated as prior assumptions. Hypotheses labels are in order of their ranking. A combination of visual and network features leads to the strongest improvement.
	}
	\vspace{-1.5em}
	\label{fig:hyptrails_results}
\end{figure}

\para{Results.}
We report the resulting Bayes factors for the comparison between expressed hypotheses and the structural baseline hypothesis (for varying hypotheses weighting factors)  in Figure~\ref{fig:hyptrails_results}; for a detailed description of this figure, please resort to the figure caption. 
The primary observation is that all investigated hypotheses improve the fit of the Markov chain model in comparison to the structural baseline hypothesis as imminent from the Bayes factors for all values of $\kappa$. 
Complementary hypotheses (e.g., navigational preference towards the core of the network) result in negative Bayes factors falling below the baseline (not shown here).
The relative ranking of hypotheses also provides interesting insights. Comparing the single hypotheses (\textit{text\_sim}, \textit{kcore} and \textit{visual}) reveals that the \textit{visual} hypothesis appears to be the most relevant, followed by the \textit{kcore} and then the \textit{text\_sim} hypothesis.
Additionally, by combining these hypotheses, we can further improve the model. Overall, a combination of all three hypotheses has the highest evidence for higher values of $\kappa$. 
Yet, the combination of belief of peripheral navigation and preference towards choosing links in the visual top and left position already provides a similarly large improvement.

\para{Summary.}
By and large, the results of this section have successfully demonstrated that we can improve models of human navigation on Wikipedia---i.e., the Markov chain model---by integrating hypotheses (beliefs) about human navigational behavior into the inference process of the models. 
In this case, we have utilized Bayesian inference to incorporate expressed hypotheses based on our insights of Section~\ref{sec:success} as priors. All hypotheses improve the model fit. Additionally, a ranking of hypotheses reveals the relative plausibility and combinations show further improvements. Next, we want to further make use of these insights to improve the well-known PageRank algorithm utilizing the Markovian framework.
\subsection{PageRank integration}
\label{sec:pagerank}

In this section, we investigate if our insights can be used for improving the well-known PageRank algorithm for Wikipedia.

\para{Weighted PageRank.}
The PageRank algorithm~\cite{page1999pagerank} is one of the most popular Web algorithms.
It computes a centrality measure---pagerank---for each node in the network, such that a node receives a high value if it has many in-coming links from other important nodes.
This can be interpreted as the probability of a random surfer in the network to land on respective page---i.e., the stationary distribution of a Markov chain model.
While for the classic PageRank algorithm, weights in the network are propagated equally through all links, we propagate weights according to the eight hypotheses that we formalized in Section~\ref{sec:HypTrails} based on our main findings.
That is, we model the landing probabilities of a random surfer that chooses the next node in the network proportional to the entries in the respective hypothesis matrix $M=(m_{ij})$.
This weighted version of PageRank implements a \emph{reasonable surfer model}~\cite{dean2010ranking}, i.e., a model where a random surfer does not choose links uniformly, but is influenced by link features such as the link position.
Formally, we compute the weighted pagerank of a node as:
\begin{equation*}
\abovedisplayskip=2pt
\belowdisplayskip=2pt
PR(j) = \frac{1-\alpha}{N} + \alpha \sum_{i \in \Gamma^{-}(j)} \frac{PR (i) \cdot m_{ij}}{Z_i}
\end{equation*} 
where $N$ is the number of nodes in the network, $\Gamma^{-}(j)$ is the set of nodes linking to the node with index $j$, $\alpha$ is the damping factor and $Z_i = \sum_{j\,' \in \Gamma^{+}(i)} m_{ij'}$ is a normalization factor for each matrix row $i$.

For evaluation, we determine the Spearman rank correlation of respective weighted PageRank with the sum of incoming transitions for a page from $D_{trans_w}$\xspace, i.e., the number of views of an article stemming from internal navigation. As a baseline, we use the classic unweighted PageRank algorithm assuming random navigation.

\para{Results.}
Table \ref{tab:pagerank} shows the resulting Spearman correlation coefficient for the different evaluated hypotheses and the unweighted PageRank as a baseline for different damping factors $\alpha$. 
We observe that most, but not all hypotheses achieve an improvement of the correlation between the aggregated transitions per target article and the pagerank values.
In particular, the \emph{kcore} and \emph{visual} hypotheses lead to improvements.
The best result can be obtained by combining these two hypotheses: the \emph{kcore\_visual} hypothesis achieves the best correlation score, which is about $0.1$ better than the baseline.
In contrast, and unexpected considering the previous results, the \emph{text\_sim} hypothesis is not able to increase the correlation, and also does not lead to improvements when combined with other hypotheses.
This might be caused by the fact that PageRank assumes an infinite random walk while people navigate in line with this hypothesis maybe only in certain phases of their Wikipedia visit, e.g., at the end of their visit, cf.~\cite{west2012human} and Section~\ref{sec:discussion}. Please note that the correlation increases with the increase of the damping factor. This is a natural behavior, since we base our analysis on Wikipedia internal transitions and higher damping factors reduce the chances of teleportation. 

Apart from that, results are mostly in line with the ranking that we obtained from the Bayesian integration of the hypothesis, cf. Section~\ref{sec:HypTrails}.
All correlation coefficients reported in this paper are strongly significant according a t-test with a null hypothesis that two datasets are uncorrelated. 
Additionally, we formally test whether the improvements of the weighted PageRank algorithms in terms of their correlation coefficients with view statistics are significant compared to the unweighted algorithm.
To that end,
we employ Steiger's one-tailed hypothesis test for assessing the difference between two paired correlations \cite{steiger1980tests}; the results reveal clear significance for all correlations when compared to the baseline.

\para{Summary.}
The results presented in this section suggest that a PageRank model with an intelligent edge weighting reflecting the users' transition behavior explains the page views stemming from internal navigation better than the standard random surfer model. The reasonable surfer implemented through the edge weighting that performs best tend to select links that are leading \emph{out of the network core to articles in the periphery and located in the top of the screen, on the left side of the screen and in the infoboxes}. 

\begin{table}[ht!]
\centering
\small
\caption{\textbf{Weighted PageRank results.} Comparison of the hypotheses specified in Section \ref{sec:HypTrails} with respect to the Spearman correlation between the transitions (aggregated per target article) and the pagerank values of the corresponding articles in $D_{wiki}$\xspace for different damping factors. 
The correlation coefficient for the unweighted PageRank (marked italic) is used as a baseline. We see that most, but not all hypotheses achieve an improvement (marked bold); the best correlation with an improvement of \raise.30ex\hbox{$\scriptstyle\mathtt{\sim}$}\,$0.1$ across the three dumping factors is achieved by the hypothesis kcore\_visual.}
\label{tab:pagerank}
\vspace{-0.5em}
\begin{tabular}{|l|l|l|l|}
\hline
\multicolumn{1}{|c|}{Damping factor $\alpha$/Hypothesis $M$ } & \multicolumn{1}{c|}{0.80} & \multicolumn{1}{c|}{0.85}         & \multicolumn{1}{c|}{0.90} \Bstrut\Tstrut \\ \hline
\Bstrut\Tstrut
\emph{baseline}                     & \multicolumn{1}{c|}{\textit{0.421}}    & \multicolumn{1}{c|}{\textit{0.428}} & \multicolumn{1}{c|}{\textit{0.436}}    \\ \tableSmallGap
kcore & \textbf{0.434} & \textbf{0.440} & \textbf{0.447} \\ \tableSmallGap %
visual & \textbf{0.507} & \textbf{0.516} & \textbf{0.526} \\ \tableSmallGap %
text\_sim & 0.400 & 0.407  & 0.415  \\ \tableSmallGap %
text\_sim+kcore & 0.407 & 0.412 & 0.417 \\  \tableSmallGap%
text\_sim+visual & \textbf{0.489} & \textbf{0.500} & \textbf{0.513}  \\ \tableSmallGap %
kcore+visual & \textbf{0.530} &  \textbf{0.538} &  \textbf{0.545} \\ \tableSmallGap %
kcore+visual+text\_sim & \textbf{0.494} & \textbf{0.505} & \textbf{0.517} \\  %
\hline
\end{tabular}
\vspace{-1.5em}
\end{table}

\section{Related Work}
Since the inception of the Web, our research community has been interested in studying human navigational click data on the Web---e.g., see \cite{huberman1998strong}.
In this line of research, a variety of models has been proposed including the well-known Markov chain model utilized in this work \cite{singer2014detecting,page1999pagerank,pirolli1999,singer2015hyptrails}, or models such as decentralized search \cite{helic2013models,dimitrov2015role} motivated by small-world navigation \cite{kleinberg2000navigation}.

Insights have been utilized to infer missing links \cite{west2015mining}, to predict break-ups of the navigation process \cite{scaria2014last}, for recommendations \cite{wang2008website}, or to improve the link structure of a website  \cite{paranjape2016improving}.
For the latter, Paranjape et al. \cite{paranjape2016improving} 
highlighted the importance of improving hyperlink structures based on their usage due to the large amount of unused links. 
Consequently, they proposed an algorithm for suggesting useful links and estimate their success based on clickthrough rates.

Based on this wide range of studies aiming at understanding human Web navigation, a series of navigational regularities, patterns and strategies have been suggested. 
For example, West and Leskovec \cite{west2009wikispeedia} found trade-offs between similarity and degree in navigational behavior suggesting different phases in user sessions, namely an exploration (orientation) and an exploitation (goal-seeking) phase \cite{helic2013models}.
Subsequent research has suggested that humans prefer to navigate between semantically similar websites \cite{west2012human,singer2013,singer2015hyptrails}, have preferences for choosing links at the beginning of pages \cite{lamprecht2016structure,paranjape2016improving,dimitrov2016visual}, and that navigational patterns exhibit regularities with respect to underlying network characteristics \cite{west2012human,paranjape2016improving,singer2015hyptrails,lamprecht2016structure}.

While, as elaborated, a large amount of studies on human navigational behavior exist and a variety of navigational hypotheses have been proposed, there still has been a lack of a systematic understanding of what makes a link successful in information networks.
To that end, we have extended previous work covered in this section by providing a study on effects of link properties on actual user transition behavior on the complete English Wikipedia utilizing large-scale data. Instead of providing a global view on clickthrough rates, we have also focused our attention on the clear setting of given a current Wikipedia article, which features best predict the popularity of links on that article. The features of interest have been motivated by previous work and our results broadly confirm previous hypotheses. Also, our analyses allow for more fine-grained insights such as navigational preference towards peripheral nodes or visual preferences on detailed screen coordinates. We have also demonstrated the importance of better understanding human navigational behavior and link popularity by successfully enriching existing Markov chain models and the PageRank algorithm with supplementary behavioral hypotheses.

Our work is also broadly connected to research studying click data in other contexts, e.g., characterizing user behavior and sybil detection in online social networks \cite{benevenuto2009characterizing, wang2013you}, improving search engine ranking functions~\cite{joachims2002optimizing,xue2004optimizing,joachims2003evaluating,agichtein2006improving,jung2007click}, and marketing and next purchase prediction~\cite{chatterjee2003modeling,bucklin2009click,bucklin2003model,montgomery2004modeling}.
However, these do not cover the specifics of large-scale information networks such as Wikipedia.

\section{Discussion}
\label{sec:discussion}

In this section, we discuss limitations and future work.

\para{No session information.} In this paper, we have focused on aggregated click counts for link transitions.
While this was perfectly suited for our task at hand, we cannot differentiate between potentially varying phases within navigation sessions as e.g., zoom-in and zoom-out phases postulated in \cite{west2012human}.
Also, we have only looked at \emph{internal navigation} while observed behavior might differ when people navigate between different platforms (e.g., Google to Wikipedia).

\para{Multiple link occurrence.}
Although the Wikipedia editor guidelines discourage placing multiple occurrences of links to the same target article within one source article, this is sometimes ignored by editors. 
Additionally, links can be repeated in the infoboxes or overview tables for improved usability.
Unfortunately, the data at hand does not disambiguate which link instance has been clicked if a link occurs multiple times.
While this has no implications for the network and semantic features as they are independent of the link position, it influences the analysis of the visual features. 
In this paper, we have decided to use the visual feature of the first occurrence of a link on the screen based on its x- and y-coordinates. 
This choice is justified by additional experiments on a restricted dataset filtering out links with multiple occurrences. 
As for the main data, a strong preference for links in the top and the left hand side of the page can be found, cf. also Section~\ref{sec:results}. 
The results for the restricted dataset confirm our main findings, but also reveal a bias: the popularity of links at the top of the page is less pronounced (but still substantial) in this variation.
For further analysis, we included the number of times a link occurs on an article as a nuisance parameter to our regression model, but this did not change the main inference.
Thus, we are confident that our main findings are robust with respect to that issue.
However, in future---preferably supplemented with more fine-grained click data---further studies on the impact of multiple link occurrences should be conducted. 

\para{Different screen resolutions.} In our analysis, we 
chose the specific screen resolution of 1920 $\times$ 1080 for deriving visual features. While this adheres to our data only containing click information from desktop users only, different users may have different visual perception of the same article.
Since this may affect presented results, future work could set out to tackle this limitation by, e.g., repeating the visual analysis for different screen resolutions. Another issue of special interest would be to complement the analysis using transitions data capturing the mobile user behavior on Wikipedia.

\para{Definition of success.} We defined success of a link as the click popularity derived from user transitions. However, other forms of link success can be studied. For example, instead of explaining transition preference of out-going links, one could modify the research question of this study and look at in-going links. Then, the question of interest would be: Given a set of in-going links to a given article, which ones are more popular than others and how can we explain them? 
Also, in this work, we have not considered the narrow textual context of a link which is tightly related to the concept of information scent \cite{pirolli2007information}. Further research can concentrate on this aspect and study the relationship between diminishing information scent \cite{hearst2009search} caused, i.e., by editors and the success of a link. Apart from that, the success of a link could also be defined in light of how much a link contributes to the different types of navigation. One could analyze navigability of the networks $D_{trans}$\xspace and $D_{trans_w}$\xspace on the edge and node level, i.e., by studying the flow in and out of a node or by removing nodes and edges.
Link success could be also examined in terms of its ability to enable and encourage further navigation and exposing the user to a more richer content or to the strongly connected component of the network.

\para{Markov chain hypotheses formulation.}
The results of Section~\ref{sec:markov} depend on how we code the hypotheses as priors.
In this paper, we hand-crafted hypotheses that are in line by our data-driven insights of Section~\ref{sec:success}.
We acknowledge that more fine-grained engineering and learning of good priors and weightings might further improve results.
Additionally, even though in different context and with different methodology, one might argue that some results of Section~\ref{sec:markov} could have been expected given our earlier findings.
Yet, the main goal here was to demonstrate the utility of incorporating assumptions about transition behavior into existing navigational models (i.e., Markov chain and PageRank) that predominantly, often have assumed uniform behavior. Additionally, obtained results helped us to further confirm our empirical insights of this paper.
Nonetheless, we hope that in the future more studies on additional datasets, e.g., on other language editions of Wikipedia, are conducted in order to enrich our understanding of human navigation and link success.

\para{Explaining and modeling heterogeneity.}
While we have accounted for basic heterogeneity by utilizing mixed-effects models and Markov chain models that both model transitions given a current start article, more fine-grained analyses in that direction are warranted in future. For example, there might be specific user groups, positions within sessions, or specific times that exhibit deviating navigation behavior.
In this direction, a detailed analysis of the random effects in the mixed-effects models might lead to better understanding of heterogeneous effects. 
Additionally, pattern mining approaches could be employed to detect interpretable subgroups of Markovian transition behavior \cite{lemmerich2016mining}. Obtained insights could then eventually be used to further advance existing models by integrating mixtures of navigational hypotheses into the modeling process \cite{mixedtrails}.

\section{Conclusions}
In this paper, we studied what makes a link successful on Wikipedia. To that end, (i) we investigated link features and their effects on link popularity utilizing descriptive analyses supplemented by mixed-effects hurdle models. Results suggest that Wikipedia users prefer to navigate to articles that are in the periphery of the Wikipedia link network, to articles that are semantically similar, and to articles that are linked at the top or at the most left-hand side of the source article.
(ii) Based on these findings, we integrated hypotheses about human navigational behavior into the well-known Markov chain model utilizing Bayesian inference. By doing so, we could improve respective model fits compared to a uniform baseline model. (iii) For further demonstrating the utility of our findings, we adapted the well-known PageRank algorithm that assumes random navigation by accounting for observed navigational preferences in a weighted variation. An evaluation of resulting pagerank values obtained by link weighting against actual view statistics revealed significant improvements over the unweighted baseline algorithm.

Our work is relevant for researchers interested in studying human navigational behavior and link structures on the Web.
Obtained insights can be utilized for providing qualitative link structures and for supplementing existing algorithms utilizing these structures to better adhere to observed transition behavior. In addition to future work directions discussed in Section~\ref{sec:discussion}, we also aim at extending our studies to different language versions of Wikipedia and newer time frames for identifying similarities and differences. Also, apart from Wikipedia, the provided methodological framework can be applied to other platform data in a straight-forward manner.

\para{Acknowledgments.} We would like to thank Denis Helic for the inspiring discussions on the PageRank section. This work was partially funded by the Austrian Science Fund project "Navigability of Decentralized Information Networks".

\clearpage
\balance
\footnotesize
\raggedright
\sloppy
\small
\bibliographystyle{abbrv}

\end{document}